# Potential Renovation of Information Search Process with the Power of Large Language Model for Healthcare

Forhan Bin Emdad, School of Information, Florida State University, Tallahassee, FL

fe21a@fsu.edu

Mohammad Ishtiaque Rahman, Department of Information Systems, University of Maryland, Baltimore County, Baltimore, MD

mrahman5@umbc.edu

## Abstract

This paper explores the development of the Six Stages of Information Search Model and its enhancement through the application of the Large Language Model (LLM) powered Information Search Processes (ISP) in healthcare. The Six Stages Model, a foundational framework in information science, outlines the sequential phases individuals undergo during information seeking: initiation, selection, exploration, formulation, collection, and presentation. Integrating LLM technology into this model significantly optimizes each stage, particularly in healthcare. LLMs enhance query interpretation, streamline information retrieval from complex medical databases, and provide contextually relevant responses, thereby improving the efficiency and accuracy of medical information searches. This fusion not only aids healthcare professionals in accessing critical data swiftly but also empowers patients with reliable and personalized health information, fostering a more informed and effective healthcare environment.

## Introduction

Artificial Intelligence (AI) has been the most discussed topic in every field, especially in healthcare research and education (Ayinde et al., 2023; Emdad et al., 2024). AI has the capacity to change the outcome of the healthcare field (Emdad et al., 2023). Advanced large language models (LLMs) can highly contribute to easing the burden of clinicians, nurses, physicians, and other health care professionals. Electronic Health Records (EHRs) integrated with AI can easily capture sound data to convert the conversation into transcription to generate correct prescriptions (Pellecchia, 2024). Furthermore, the information generated from the LLMs with appropriate prompts can quench the thirst of knowledge-seeking patients and clinicians. Even LLMs have successfully passed the US MLE exam, proving the efficiency of LLMS in the medical field (Mbakwe et al., 2023). This process of information searching and gathering can bridge a connection between the patients and clinicians. Despite the different biases and incorrect information provided by the LLMs, general information-seeking patients roam into LLMs to gather their desired information. However, sometimes patients get satisfaction from the search results of LLMs, sometimes patients are not satisfied. The reason for not finding appropriate search results can be due to nonstructured prompts. However, the objective of this study is to build a process of searching desired medical-related information search with the Information Search Process (ISP). Therefore, this study has the following objectives:

1) Development of ISP theory over the years and designing a unified ISP theory powered by AI.
2) Implementation of the proposed unified ISP theory in the healthcare field.

## Background

Information systems development is focused mostly on the system's representation rather than focusing on the users' perspective, information needs, and problems. Users' cognitive perspectives are often missing while developing information systems. Researchers were aware of this phenomenon, they were identifying ways of understanding the information behavior of users (C. C. Kuhlthau, n.d.). Still, the impact of users' feelings was not taken into consideration by the researchers. The information Search Process (ISP) model focuses immensely on the user's perspective and the role of affect in information-seeking behavior. The ISP model consists of six sequential major stages: task initiation, selection, exploration, focus formulation, collection, and presentation. Six stages represent the cognitive feelings of users. This model is inspired by Kelly's construct theory. Although, the main focus has been placed on uncertainty in ISP which is missing in Kelly's construct theory (Kelly, 1963). Kuhlthau developed the ISP model by conducting two series of empirical qualitative and quantitative studies over a decade on secondary school students about library use. The summary of the output of this study indicated identifying the uncertainty, confusion, and frustration in the thought process about the problem statement in the early stage of the user's point of view of the ISP.

In short, the ISP model comprises of user's perspective in information seeking as a set of thoughts, feelings, and actions. The major factor in this process is uncertainty identified in the early stages, which converts into confidence with a clear concept of the topic over time. In this context, the goal of ISP is not only to gather information concerning the user's information seeking but also to use the acquired information to solve the existing problem. To accomplish this goal, ISP consists of six stages considering information seeking from the user's perspective.

The first stage of ISP is "initiation", users identify potential knowledge gaps and information needs. In this stage, users feel uncertainty, and confusion, and realize the problem. "Selection" is the second stage of ISP where users identify the topic that needs to be analyzed by gathering information. Uncertainty about the information leads to an information search. Search becomes successful if the selections are made correctly. Actions are taken based on the selection stage. Uncertainty increases when the selection process is delayed because of the confusion of the topic. The next stage is "Exploration" where the confusion and frustration of users increases more. In this stage, users start identifying information about the topic. User uncertainty rises because the topic is unclear. Information sources can differ along the way causing frustration during this process. This stage is considered the most difficult stage in the ISP model. The fourth stage of this model is "Formulation" where uncertainty starts to decrease, and confidence increases. In this stage, users identify the information required for the desired aspect of the topic. The search process aligns with the assumption. Acquired insights bolster the focus and lead toward clear-cut information on the topic. This stage makes users more confident and grows the sense of clarity among users. The fifth stage of the ISP model is the collection. In this stage, users build a genuine interaction with information systems. Users gather information based on the topic which increases the focus on the topic. In this phase, general information about the topic is no longer required as the users can delve into a deeper concept of the topic. Users feel confident as a deep sense of the topic has been gained and uncertainty diminishes. The final stage of the ISP model is known as "Presentation". In this stage, users feel satisfied if the information search has been fruitful otherwise users feel disappointed. The user's goal always remains that the final information can solve the existing problem or complete the task or accomplish the goal (C. C. Kuhlthau, n.d.). Figure 1 shows the six stages of the ISP model with the change in user satisfaction.

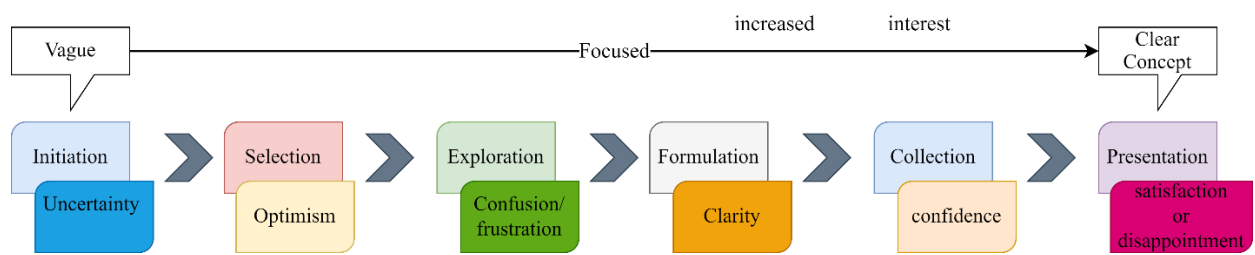

Figure 1: Initial Model of ISP with six stages

**Development of ISP over time**

The ISP model was developed by Kuhlthau as a part of the doctoral thesis defense. Later, her research was converted into a paper in 1988. Kuhlthau conducted a longitudinal study on high school students for over 4 years from the perspective of the information search process in libraries. This study was extended when the students went to college. Significant changes were noticed while the results of the two studies were compared in three areas: assignments, formulation, and procedure for information gathering (C. C. Kuhlthau, 1988).

*ISP: Information seeking from the user's perspective*

Kuhlthau's work expanded by adding the results of three more experiments. The third experiment included a large scale of high school students and library users. In the fourth experiment, high, middle, and low-grade achieving high school students were included as the population to observe whether they have similar experiences. In the fifth study, many library users were included to validate the results of the ISP model. In this study, the outcome presented differences among the perspectives of the groups. Public library users seem to show more confidence than academic library users. Interest in the topic did not increase as the search progressed, rather the topic changed with the collection of new information. Users also showed a sense of tolerance in the early stage of uncertainty. Users were responsible for showing ownership of the process. Again, there were differences in the perception of tasks among the users. Users started gathering information from the early stages of the search process. Users gathered information in almost every stage while the ISP model suggests that the exploration and formulation stage is dedicated to building user's thoughts about the topic. As a result, users presented information about the topic without focus and clarity. The presented information on the topic reflected the lack of perspective of the author rather it seemed to borrow a framework from another's perspective (C. C. Kuhlthau, 1991).

### *Introduction of zones of interventions for librarians in the ISP model*

Kuhlthau extended her findings of the study by introducing the zone intervention concept. This study demonstrated the identification of zones where intervention can be made in identifying the topic. In this context of students' information search process, librarians should involve themselves in the process of intervention. In this way, zone intervention can be helpful for the information-seeking of students. Five zones of intervention have been identified in this study. Zones of interventions are determined by the problems. Librarians should be aware of the problem in four areas: task, interest, time, and availability. Students' problems and processes can determine the level of mediation. Organizers are considered the first level of mediation. Organizers play an important role in making the collection phase self-serviced. The second level is known as the locator. The locator helps the users to identify the information sources for specific queries of the users. The third level of mediation is the identifier which is linked with zone three. The fourth level of mediation is known as an advisor. Advisors can identify the patterns of search in the information sources. The fifth level of mediation is the counselor which is linked with zone five. Counselors not only identify the information sources but also address the constructive process for learning the specific topic. There are different levels of mediators in library and university settings. The organizer is the first level, and the lecturer can be the second level of mediator related to zone two. The third level of mediator is the instructor who can provide relevant courses for instruction. The fourth level of mediator is the tutor who can help with strategies and course integration. The fifth level of mediator is the counselor related to zone five dealing with process instruction. Strategies of process intervention need further attention to understand the zones of intervention. Five strategies have been identified for process intervention: collaborating, continuing, converting, charting, and composing. Collaborating in the information search process is very important as this is not an isolated process. Librarians can be collaborators in the intervention process. This process involves brainstorming, delegation, networking, and integrating information. Continuing intervention deals with resolving the series of information problems. Conversion is conducted by the counselor who guides the students by suggesting strategies for the process. The counselor provides ways of narrowing the topic to the students so that students do not have to delve into the broader topic. The next strategy is a charting intervention which presents an overview of all the gathered information in a simplistic way. The final strategy in the process intervention is composing which helps users to formulate thoughts and develop a sense of constructive process in the information search. Counselors can help users implement a composing strategy in the search process (C. Kuhlthau, 1994).

*Learning in digital libraries with the ISP approach*

The evolution process of ISP took a new turn when the digital library concept emerged to the surface. Kuhlthau adopted that concept into the model. Schools and libraries were adopting the changes of the information age. Open communities were established between institutions with the help of online computers. Educators felt the need for a new constructive approach to learning. Constructivists took the approach of transforming existing books into digital libraries. Teachers and librarians played a vital role in the transformation according to the constructivist approach. Teachers and librarians made a collaborative effort in this process of creating digital libraries. The ISP was established based on five strategies. These strategies were adaptive for digital libraries. Creating learning environments in digital libraries was the main focus of the researchers. Learning is the process of changing uncertainty to a clear understanding through information seeking. The constructive process of learning consists of four phases: acting and reflecting, feeling and formulating, predicting and choosing, and interpreting and creating. Librarians can guide users to act and reflect on perspective before starting to read. "Acting and reflecting" should begin in the early stages of the ISP model. "Feeling and Formulating" should be applied to the search process to have the user's view on the topic after information gathering to remove uncertainty and understand the topic better. Librarians can help users in better formulating. "Predicting and Choosing" enables users to choose the areas where users can delve into for profound understanding through the information search process. "Interpreting and Creating" enables users to create information beyond the provided information which is the main goal of the ISP model (C. C. Kuhlthau, 1997).

*Revisiting ISP for Coping with the Digital Age*

Kuhlthau realized the problem that information environments do not remain static but change with the use of technology. Kuhlthau conducted a study to revisit the ISP model coping with the modern age. Digital library environment creation was an advancement in the process where it has been integrated with the ISP model. Information retrieval also played an important role for students and other users in the information search process. This model has been an effective way for student's information seeking from libraries. In addition, this model provided guidance on where to establish intervention in the search process. Diagnostic characteristics of the ISP model have played an important role in the modern age because users have availability and access to a huge amount of information. Students tend to focus on the final product without investigating the background knowledge. The exploration, focus, and formulation phase of the ISP model guides the students to have thoughts in the early stages of the search process. The ISP model helps users develop a data collection framework and ways of turning information into knowledge considering the value-added approach. The results of this study are backed up with a constructive view. Feelings can be considered as the key factor while information accumulation of certain topics turns to clear understandable information used for solving problems. Finally, the results of this study indicate that the ISP model still remains relevant in the age of technology (C. Kuhlthau et al., 2008). Figure 2 demonstrates an integrated view of the ISP model development over the years.

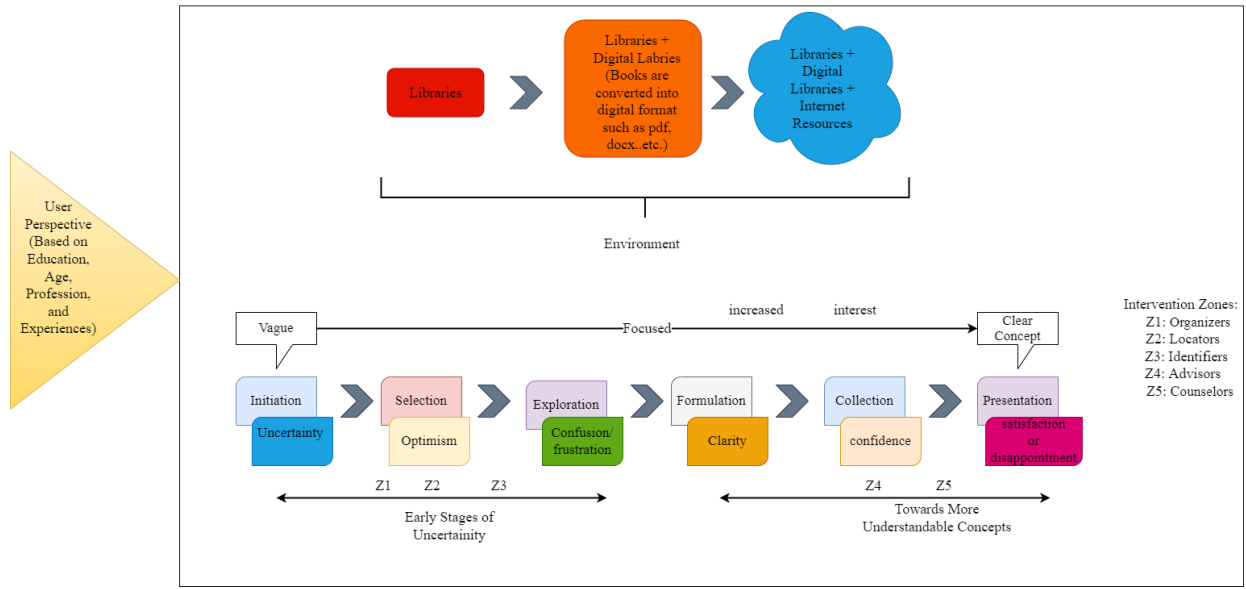

Figure 2: Final ISP model

**Factors Influencing ISP**

There are a few factors that influenced the ISP model: uncertainty, confidence, user perspective, interventions, and technology. ISP model was based on the process where the user's feeling of uncertainty turns into confidence. At the initial stages of the ISP model users possess unclear concepts of the topic. As a result, uncertainty becomes prevalent in the initial stage of the ISP model. As the search progresses into the final stages, the topic becomes more understandable, and confidence increases. The user's perspective is another important factor where the user's experience and education play a role in influencing the ISP model. Interventions in the different stages of ISP have influenced the ISP model. Librarians and teachers have played a positive role in influencing the ISP model through intervention. Lastly, technology is the most important factor in influencing information search. As technology is advancing, the library has turned into a digital library, and the paper-based book has turned into digital books. ISP model has adopted the change and encouraged to creation of digital libraries. ISP model has evolved with time to be relevant in the modern age.

**Implementation of ISP in Different Domains**

*Children's web searching behavior using the ISP approach*

Children's search behavior also utilizes the ISP framework in online search engines. Children's web searching experiment included two major search engines for this purpose. Broch (2000)'s study also discussed some difficulties faced by the children during the search process. Researchers suggested that children have an inclination towards better graphical interfaces for searching. Usually, there is a thinking that children of modern age are tech savvy. However, the results of the study have shown a low success rate in using the web for searching among children. As a solution for this existing problem, teachers and librarians can work collaboratively to help children acquire internet searching skills (Broch, 2000).

*Exploring Collaborative Information Behavior (CIB) with ISP in the group-based educational setting*

Kuhlthau's study was conducted depending on individuals' experience in the information search process while developing the ISP model. This ISP model approach was applied to find the Collaborative Information Behavior (CIB) in a group educational setting. The social and collaborative factor is essential for identifying user experiences, user roles, and interactions while solving a problem. A study by Hyldegård (2006) was conducted with two groups of graduate students in library and information science. Group members initially have a weak focus, but social factors have a major impact on them. Some students in the group felt disappointed and frustrated at the end. Although, a mismatch was observed in the emotions, feelings, and ambitions among the group members. Lack of time created an influence on the feelings of the students and groups in the study. The results of this study suggested that sometimes uncertainty does not turn into confidence. In this case, the negative feelings did not change into positive feelings which was the turning point ISP model. Although, the collaborative approach helped group members to better understand the topic and influence the feelings. Finally, social and collaborative factors had an effect in the stages of the ISP model but feelings varied among the individual group members (Hyldegård, 2006).

*Information Search Process of Lawyers: A Call for 'JUST FOR ME' Information Services*

Another study by Kuhlthau & Tama has presented that a user's affective experience has some critical influence on information-seeking behavior. However, affective experience shapes the construction of new knowledge. Previous information systems were designed to help information seeking and information collection rather than implementation to accomplish the user's goal. Still, there is a lack of information search processes to complete a complex task. This study focuses on information workers' perspectives on the information search process to complete their complex tasks. This study is a qualitative longitudinal study conducted with eight practicing lawyers, four males, and four females. Participants' experience varied from six to ten years in the law farms. Data was collected in the form of semi-structured interviews. Complex tasks for the lawyers include preparing for trial, fact gathering, and solving the problem through trial. Formulation and constructing new thoughts require a lot of thinking for the lawyers while completing the complex task. Lawyers have to apply different approaches to accomplish their goals. The summary of the findings suggests that lawyers felt uncertainty at the beginning of the search process, but uncertainty turned into confidence with the progress of the ISP model. Although, experts responded differently to the uncertainty than novice lawyers. This result reflects that experience plays a crucial role in information search. Lawyers mentioned that there is a gap in generations as the materials have turned from paper-based to electronic (C. C. Kuhlthau & Tama, 2001).

*Immigrants' information needs: their role in the absorption process*

Immigrants have specific needs for information as they travel from one place to another in search of better living and employment leaving behind the harsh reality of war, deprivation of food, and poverty. Another Kuhlthau's study investigated the information needs and satisfaction of immigrants aligning with the ISP model. Data were collected through a qualitative study including thirteen interviews with immigrant families emigrating from North America to Israel. The findings of this study suggest that immigrants require a lot of information at the beginning stage of the immigration process. Immigrants feel a need for critical information related to health, schooling, housing, and banking. Although, there is a lack of information in English regarding these essential issues. Moreover, existing information is insufficient to answer specific questions about immigrants. As a result, Immigrants remain unsatisfied after the information search process because of the challenges of retrieving information and information scarcity. Kuhlthau's ISP model suggests that information seeking from information search should be conducted not only to collect and gather information but also to accomplish a task or goal (C. C. Kuhlthau, n.d.). During this study, this purpose is not served fully (Snunith Shoham, n.d.).

**Proposed Unified ISP Powered with AI**

The unified information search process leverages the capabilities of large language models (LLMs) to streamline and enhance the efficiency of retrieving relevant information. By integrating sophisticated natural language processing (NLP) techniques (GuYu et al., 2021), Kuhthau's ISP will enable a seamless interaction between users and vast data repositories. This process will involve the LLM understanding and interpreting domain-specific user queries in a highly nuanced manner, facilitating the extraction of precise and contextually appropriate information across diverse datasets (Chen et al., 2020). Better prompting techniques will be essential to conduct the interactions between the users and LLMs. The LLM model's ability to comprehend and generate human-like text allows it to bridge gaps between varied information sources, presenting coherent and consolidated search results. As a result, users benefit from a more intuitive and comprehensive search experience, where the complexity of data retrieval is significantly reduced, and the relevance of the information provided is markedly increased. This unified search mechanism not only enhances the user experience but also sets a new standard for information accessibility in the digital age. Figure 3 provides the visualization of LLM powered ISP model.

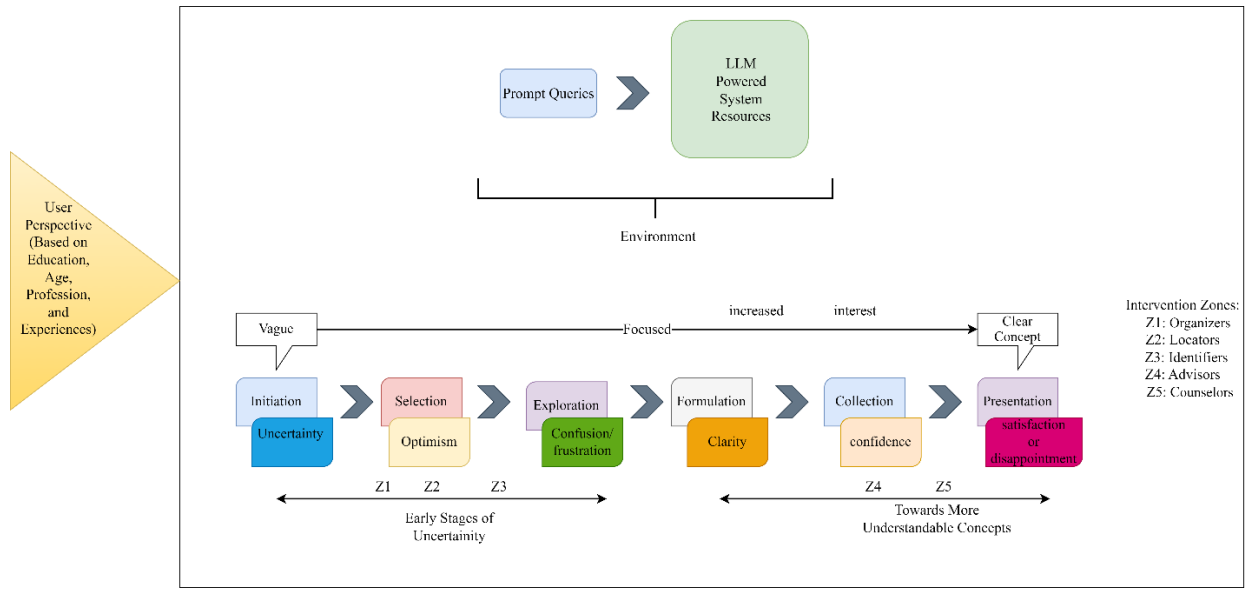

Figure 3: Proposed Unified LLM Powered ISP Model

**Implementation of ISP in Healthcare**

Implementing an information search process powered by large language models (LLMs) in healthcare revolutionizes the way medical professionals and patients access critical data. By harnessing the advanced natural language processing capabilities of LLMs, this system can interpret complex medical queries and retrieve relevant information from vast and diverse medical databases with remarkable accuracy. The process begins with the LLM understanding the nuances of the user's question, whether it pertains to consumer health, symptoms, treatments, drug interactions, or recent research findings (Zhang, 2012). It then navigates through extensive healthcare records, journals, and clinical guidelines to provide precise and contextually appropriate answers. This reduces the time healthcare providers spend searching for information, allowing them to focus more on patient care. Additionally, the system can assist patients in understanding their conditions and treatment options, promoting informed decision-making and better health outcomes. The integration of LLMs in healthcare search processes not only enhances the efficiency and accuracy of information retrieval but also supports a more personalized and responsive healthcare system.

**Conclusion**

In conclusion, the integration of an information search process powered by large language models (LLMs) in healthcare represents a significant advancement in medical information retrieval and accessibility. This technology enhances the ability of healthcare professionals to quickly and accurately access vital information, thereby improving the quality and efficiency of patient care. Patients also benefit from more informed and personalized healthcare experiences, as LLMs can interpret and provide relevant information tailored to their specific needs and queries. By reducing the time and effort required to find accurate medical information, LLMs enable a more responsive and effective healthcare system. As this technology continues to evolve, its potential to transform healthcare delivery and outcomes will only grow, highlighting the critical role of advanced AI in the future of medicine.